\documentclass[letterpaper]{article}
\usepackage{aaai2027} 
\nocopyright
\usepackage[hyphens]{url} 
\usepackage{graphicx} 
\usepackage{natbib} 
\usepackage{caption} 
\usepackage{subcaption}
\usepackage{xcolor}
\usepackage{mdframed}
\usepackage{amsmath} 
\usepackage{booktabs}
\usepackage{amsfonts} 
\usepackage{algorithm}
\usepackage[noend]{algpseudocode}

\begin{document}

\title{SeqFeed: Improving LLM-Based RTL Code Generation with Sequential Behavior Feedback}

\author{
    Yuxin Du\equalcontrib\textsuperscript{\rm 1},
    Juxin Niu\equalcontrib\textsuperscript{\rm 1},
    Tao Hu\textsuperscript{\rm 1},\\
    Xi Wang\textsuperscript{\rm 2,\rm 3},
    Zhe Jiang\textsuperscript{\rm 2,\rm 3},
    Nan Guan\textsuperscript{\rm 1}
}
\affiliations {
    \textsuperscript{\rm 1}City University of Hong Kong,\\
    \textsuperscript{\rm 2}Southeast University, China,\\
    \textsuperscript{\rm 3}National Center of Technology Innovation for EDA, China\\
    yuxindu8-c@my.cityu.edu.hk,
    juxin.niu@my.cityu.edu.hk,
    hu.tao@cityu.edu.hk,\\
    xi.wang@seu.edu.cn,
    101013615@seu.edu.cn,
    nanguan@cityu.edu.hk
}
\maketitle

\begin{abstract}

RTL code generation is a critical stage in hardware design, and the emergence of agentic systems offers new opportunities to automate this process. To generate correct RTL code, agents must understand sequential behavior, including how signals evolve and propagate over multiple clock cycles. However, effectively conveying such temporal information to agents remains a significant challenge.
RTL source code does not directly reveal cycle-by-cycle signal evolution in a concrete execution, while full simulation waveforms are too large and contain too much irrelevant information for effective use by LLMs.
To address these limitations, we study how human engineers reason about sequential behavior and identify three requirements for effective execution feedback: it should be event-addressable, dependency-traceable, and incrementally expandable.
Guided by these requirements, we propose SeqFeed, a framework with two complementary components: (1) SeQuery, an SQL-like waveform query language that enables agents to anchor queries to semantic events and sample signal values at relative time points, and (2) SeGraph, a dependency graph that tracks signal propagation across clock cycles.
Experimental results across multiple LLMs demonstrate the effectiveness of SeqFeed in improving pass rates. SeQuery and SeGraph are each effective independently and provide complementary benefits when used together. Further analysis indicates that SeqFeed shifts agent reasoning from speculation-driven conjecture to evidence-based analysis.

\end{abstract}

\section{Introduction}

The use of LLMs for complex tasks is shifting from static, single-turn prompting toward agentic problem solving, where an LLM agent autonomously invokes tools, reasons through multi-step trajectories, and iteratively refines its output through self-correction. This approach has been adopted across application domains including software engineering, scientific discovery, and task automation~\cite{wang2024survey,yao2023react,madaan2023selfrefine}. 
A central insight is that an agent's ability to improve iteratively depends on whether its environment provides feedback that is both \textit{informative} and \textit{actionable}. Such feedback should reveal how the current output deviates from the target and provide concrete guidance for correction~\cite{gou2024critic}. 

RTL code generation is a critical stage in hardware design, translating functional specifications into synthesizable hardware description languages such as Verilog. Traditionally, this process has required substantial manual effort and remains susceptible to subtle logic errors~\cite{abdollahi2024hardware}. Although LLMs have shown promising results in automating RTL code generation~\cite{liu2023verilogeval,tsai2024rtlfixer,chang2024data}, their integration into agentic workflows remains underexplored.

Producing correct RTL code requires an agent to understand its sequential behavior, including how signals evolve and propagate across clock cycles. To iteratively improve the generated RTL, the agent needs effective feedback that helps it understand such behavior and identify appropriate code revisions. 
However, providing this feedback in an effective way remains challenging. 
The two most direct sources of sequential information are RTL source code and simulation waveforms. These two sources provide complementary views of sequential behavior, but each has limitations when used for agents.

\begin{figure}
  \centering
  \includegraphics[width=.9\columnwidth]{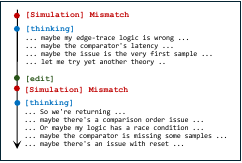}
  \caption{Without cycle-level evidence, an agent may repeatedly revise RTL based on unsupported hypotheses.}
  \label{fig:challenge_example}
\end{figure}

\textit{RTL source code} describes the design logic and the rules governing state transitions. However, it does not directly reveal which path conditions hold or how signals evolve in a specific execution~\cite{yao2025arspautomatedrepairverilog}. A simulation mismatch indicates that the generated RTL does not satisfy the intended behavior, but it does not explain which sequential behavior caused the mismatch. As illustrated in Figure~\ref{fig:challenge_example}, without targeted cycle-level execution evidence, the agent may generate multiple competing hypotheses that are all plausible from the source code. It must then select one hypothesis, revise the RTL, and re-simulate the design. If the mismatch persists, the agent repeats the same process with another unsupported hypothesis, resulting in an inefficient cycle of conjecture, revision, and re-simulation.

\textit{Simulation waveforms} complement RTL source code by providing concrete execution evidence, including cycle-level signal values and state transitions. Such evidence can help distinguish among hypotheses that cannot be resolved from the source code and the simulation result alone. Recent work has demonstrated the potential of waveform feedback for LLM-based RTL generation~\cite{ho2025verilogcoder}. However, full waveforms often contain many signals and cycles unrelated to the agent's current reasoning, overwhelming the context window with unnecessary information. Besides, raw waveforms show what occurred during execution but do not directly connect the observed behavior to the responsible RTL logic or indicate how the code should be revised.


Another possible feedback mechanism is to instrument the RTL with diagnostic constructs such as \verb|$display|. However, the instrumentation must be specified before simulation, and modifying the observed signals or trigger conditions requires rerunning the simulation. Because simulating a complex design can take minutes or even hours, repeated instrumentation and re-simulation can substantially slow the agent's iterative refinement process.

Motivated by these challenges, we make the following three contributions:

\begin{itemize}
    \item We derive three design requirements by examining how human engineers jointly use RTL source code and simulation waveforms to diagnose sequential behavior.
    
    \item We propose \textit{SeqFeed}, which includes \textit{SeQuery} for event-based waveform querying and \textit{SeGraph} for cycle-aware dependency tracing.
    
    \item We evaluate \textit{SeqFeed} on RTL code generation benchmarks across multiple LLMs, demonstrating that it improves success rates and enables more evidence-based diagnosis and refinement.
\end{itemize}

\section{Motivation}

Agentic RTL generation requires effective feedback about sequential behavior.
To design such feedback, we draw on common practices used by hardware engineers to interpret sequential executions and distill three recurring patterns.

\textit{First, engineers anchor their inspection on events.}
Rather than scanning a waveform cycle by cycle, they typically begin by locating a specific event, such as the completion of a handshake, and then inspect its surrounding cycles.
The event provides a reference point, while the surrounding window provides context.
\textit{Second, engineers trace dependencies backward.}
After identifying a signal whose behavior requires explanation, engineers trace the upstream assignments, enabling conditions, and earlier signal values that determine its value.
They follow these dependencies backward across clock cycles, focusing the inspection on relevant signals at progressively earlier cycles.
\textit{Third, engineers refine their inspection scope iteratively.}
Engineers seldom determine all relevant signals and cycles in advance.
Instead, each observation informs a revised hypothesis, which suggests additional signals or cycles to inspect.
The inspection scope thus expands progressively as their understanding develops.

Together, these patterns determine \emph{where} to look, \emph{what} is relevant, and \emph{how much} information to examine.
From them, we derive three requirements for a sequential feedback mechanism for LLM agents:

\begin{itemize}
  \item \textit{Event-addressable.}
  The mechanism should allow an agent to locate cycles in which a specified signal event occurs and inspect selected signals within a surrounding cycle window, without scanning the entire waveform.

  \item \textit{Dependency-traceable.}
  Given a signal at a specific cycle, the mechanism should identify the RTL assignments, enabling conditions, and earlier signal values that determine its observed value.

  \item \textit{Iteratively-queryable.}
  The mechanism should allow an agent to issue successive queries based on previously obtained evidence, without requiring the full inspection process to be specified in advance.
\end{itemize}

\begin{figure*}[t]
    \centering
    \includegraphics[width=\textwidth]{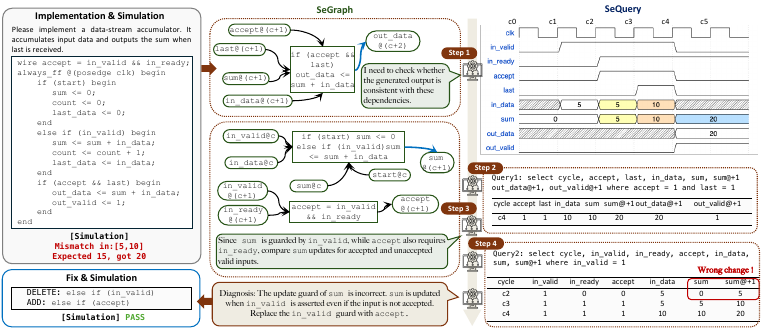}
    \caption{Overview of SeqFeed.}
    \label{fig:overview}
\end{figure*}

\section{SeqFeed}

\subsection{Overview}

SeqFeed provides effective sequential feedback to agents through \textit{SeQuery} and \textit{SeGraph}:

\begin{itemize}
  \item \textit{SeQuery} is a waveform query language. When an agent needs to answer questions such as ``at which cycle did a handshake succeed, and what were the signal values around it,'' it writes a query targeting the relevant event. Starting from a simple condition, the agent refines or expands the query as its debugging hypothesis evolves.

  \item \textit{SeGraph} is a dependency graph extracted from RTL source code. Given a signal of interest, it returns an upstream subgraph organized by clock cycle: which assignments, conditions, and upstream signals contributed, and over how many cycles. The agent follows this dependency chain backward to the root cause.
\end{itemize}

SeqFeed does not prescribe a fixed invocation order between the two; the agent decides autonomously when to query the waveform, when to trace dependencies, and how to interleave them as hypotheses evolve.
Figure~\ref{fig:overview} illustrates this process with a concrete example.

Consider a data-stream accumulator. The agent generates an initial RTL implementation and simulates it, finding a mismatch: input $[5,10]$ yields $20$ instead of the expected $15$. To locate the error, the agent first traces \verb|out_data| backward through SeGraph and learns that \verb|out_data = sum + in_data|. It then issues a SeQuery at the cycle where the final input is accepted: \verb|sum = 10|, \verb|in_data = 10|, so \verb|out_data = 20| is consistent with that expression. Logically, however, \verb|sum| should be $5$ before processing the last input of $10$, which means \verb|sum| was already accumulated incorrectly earlier. Suspecting an error in how \verb|sum| is updated, the agent traces \verb|sum|'s dependencies through SeGraph and discovers that the gating condition is \verb|in_valid| rather than \verb|accept|. After confirming this finding with a second SeQuery, the agent corrects the gating condition to \verb|accept|, regenerates the code, and the simulation passes.

The next two subsections detail \textit{SeQuery} and \textit{SeGraph}, respectively, while the final subsection discusses their design trade-offs between tool completeness and LLM usability.

\subsection{SeQuery}

\begin{table*}[t]
\centering
\caption{SeQuery examples across five sequential checking scenarios.}
\label{tab:sequery-examples}
\begin{tabular}{p{0.45\textwidth} p{0.5\textwidth}}
\toprule
\textbf{Natural language description} & \textbf{Query} \\
\midrule
{\small\textit{Detect data latch errors.} When a transfer is accepted at cycle $t$, the latched data at $t+1$ must match the input at $t$.}
& {\small$\text{select}\ \text{cycle}, \text{datain}, \text{data\_latched}@+1$ \newline
  $\text{where}\ \text{valid} = 1 \land \text{ready} = 1 \land \text{data\_latched}@+1 \neq \text{datain}$} \\
\midrule
{\small\textit{Check data stability under backpressure.} When $\text{valid}=1$ and $\text{ready}=0$, the source must hold both $\text{valid}$ and $\text{data}$ stable until the transfer completes.}
& {\small$\text{select}\ \text{cycle}, \text{valid}, \text{ready}, \text{valid}@+1, \text{data}, \text{data}@+1$ \newline
  $\text{where}\ \text{valid} = 1 \land \text{ready} = 0 \land (\text{valid}@+1 \neq 1 \lor \text{data}@+1 \neq \text{data})$} \\
\midrule
{\small\textit{Check a fixed-latency pipeline.} An input accepted at $t$ should produce valid output at $t+3$.}
& {\small$\text{select}\ \text{cycle}, \text{in\_data}, \text{out\_valid}@+3, \text{out\_data}@+3$ \newline
  $\text{where}\ \text{in\_valid} = 1 \land \text{in\_ready} = 1 \land (\text{out\_valid}@+3 \neq 1 \lor \text{out\_data}@+3 \neq \text{in\_data})$} \\
\midrule
{\small\textit{Check FSM state transitions.} A transition from state $0$ to $1$ is allowed only when $\text{start}=1$.}
& {\small$\text{select}\ \text{cycle}, \text{state}, \text{state}@+1, \text{start}$ \newline
  $\text{where}\ \text{state} = 0 \land \text{state}@+1 = 1 \land \text{start} \neq 1$} \\
\midrule
{\small\textit{Detect pipeline bubble stalls.} A pipeline expected to produce every cycle must not idle for 3 or more consecutive cycles.}
& {\small$\text{select}\ \text{cycle}, \text{out\_valid}, \text{out\_valid}@+1, \text{out\_valid}@+2$ \newline
  $\text{where}\ \text{out\_valid} = 0 \land \text{out\_valid}@+1 = 0 \land \text{out\_valid}@+2 = 0$} \\
\bottomrule
\end{tabular}
\end{table*}

To understand how a SeQuery maps to its result, consider Query~1 in Figure~\ref{fig:overview}.
The $\text{select}$ clause specifies which signal values to retrieve for each matching cycle.
The $\text{where}$ clause restricts the result to cycles where $\text{accept}=1$ and $\text{last}=1$ hold simultaneously.
The $s@+1$ suffix samples a signal one cycle after the matched cycle.
The result is a table with one row per matched cycle and one column per $\text{select}$ item.

Three core primitives emerge from this example.
First, $\text{select}$ defines the columns of the result table: for each matching cycle, the user specifies which signals and which offsets to observe.
Second, $\text{where}$ is a Boolean condition over signal values that filters anchor cycles; it supports comparisons, edge-detection functions, and and/or/not combinations.
Third, the $s@k$ notation samples signal $s$ at an offset of $k$ cycles relative to the anchor, with $k>0$ looking forward and $k<0$ looking backward.
These primitives compose: more complex checks are assembled by adding conditions in $\text{where}$ and offset samples in $\text{select}$.

Table~\ref{tab:sequery-examples} presents further examples of these primitives applied to data latching, backpressure, fixed-latency pipelines, FSM state transitions, and pipeline liveness.

Formally, a query is defined by the following grammar:
\[
\begin{aligned}
Q \;::=&\; \text{select}\ I\;(\text{,}\ I)^{*}\ \text{where}\ P \\[2pt]
I \;::=&\; \text{cycle} \mid s@k \mid \text{min}(\text{cycle}) \mid \text{max}(\text{cycle}) \mid \text{count}(*) \\[2pt]
P \;::=&\; E \bowtie E \mid \text{rise}(s) \mid \text{fall}(s) \mid \text{change}(s) \\
        &\mid \text{not}\ P \mid (P) \mid P \land P \mid P \lor P \\[2pt]
E \;::=&\; \text{cycle} \mid s@k \mid c
\end{aligned}
\]

In the grammar above, $s$ denotes a signal name, $k \in \mathbb{Z}$ is an integer cycle offset, $c$ is a constant, and $\bowtie \in \{=, \neq, <, \leq, >, \geq\}$ is a comparison operator.
Evaluation proceeds under an externally specified clock domain and active edge.
The evaluator enumerates all clock cycles within the query range, treating each cycle in turn as the anchor for $\text{where}$ predicate evaluation.
Signal values are sampled before the active clock edge; $s@k$ denotes the sampled value of signal $s$ at $k$ cycles relative to the current anchor.
For brevity, a bare signal name defaults to $s@0$.
Comparisons operate on four-state logic values ($0$, $1$, $X$, $Z$).
Among the edge-detection functions, $\text{change}(s)$ tests whether adjacent samples differ in any way, while $\text{rise}(s)$ and $\text{fall}(s)$ detect only strict single-bit $0 \to 1$ and $1 \to 0$ transitions. Aggregate functions such as $\text{min}(\text{cycle})$ and $\text{count}(*)$ let the agent summarize the set of matching cycles, for instance, identifying the first or last occurrence of an event.

\subsection{SeGraph}

To understand what SeGraph provides, consider the \verb|out_data| dependency chain in Figure~\ref{fig:overview}.
Signals \verb|accept| and \verb|last| feed into the condition node \verb|accept && last|, which gates the computation node \verb|sum + in_data|; the result is assigned to \verb|out_data|.
This chain is organized by clock cycle rather than by RTL source line order.
The assignment of \verb|accept|, the test of \verb|last|, and the computation of \verb|sum| appear in separate code locations, yet every node carries a cycle annotation that makes cross-cycle value propagation visually explicit.
This cycle-based organization contrasts with conventional dependency graphs and AST-based representations, which follow code structure.

Three structural properties emerge from this example.

\begin{itemize}
  \item \textit{Node.}
  SeGraph uses two logical node types internally. A \emph{signal node} represents an RTL signal. A \emph{statement node} captures the RTL code that produces one or more signals, including conditions, computations, and control structures (\verb|if|/\verb|case|).
  When presented to the agent, each statement node is expanded: conditions become condition nodes (diamonds in Figure~\ref{fig:overview}), computations become computation nodes (rectangles), and both connect to their signal operands (rounded rectangles).
  This expansion yields the three visual forms shown in the figure.

  \item \textit{Edge.}
  All dependencies form alternating chains of signal and statement nodes.
  A statement node reads its input signals, evaluates conditions, performs computations, and assigns the results to output signals.
  Both value operands (e.g., \verb|sum| and \verb|in_data| feeding the addition in the \verb|out_data| chain) and control conditions (e.g., \verb|accept| and \verb|last| forming the gating condition) are treated as statement inputs; the complete causal context is thus preserved within the graph.

  \item \textit{Cycle-based organization.}
  SeGraph groups nodes by the clock cycle to which they belong.
  A combinational statement and its input and output signals all reside within a single cycle group, as in the \verb|out_data| chain of Figure~\ref{fig:overview}.
  A clocked statement marks a cycle transition: its input signals belong to cycle $n$ and its output signals to cycle $n+1$.
  In the \verb|sum| update subgraph of Figure~\ref{fig:overview}, \verb|sum| and \verb|in_data| combine in the current cycle and the result becomes \verb|sum@+1| in the next, crossing the cycle boundary.
  A multi-cycle dependency therefore appears as a path through successive cycle groups, making data propagation across clock ticks directly traceable.
\end{itemize}

These structural properties are formalized in a static graph $G = (V, E)$, built once from the RTL design.
$V$ partitions into signal nodes $V_{\text{sig}}$ and statement nodes $V_{\text{stmt}}$; edges alternate between the two sets, so $E \subseteq (V_{\text{sig}} \times V_{\text{stmt}}) \cup (V_{\text{stmt}} \times V_{\text{sig}})$.
Given a target signal $s \in V_{\text{sig}}$, SeGraph extracts a subgraph by backward traversal: from a signal to its producing statements, from a statement to its input signals.
Traversal terminates at primary inputs, undefined signals, visited nodes, or a depth limit.
Symbolic unfolding is applied during traversal: combinational statements preserve the current cycle, while clocked statements advance it by one.
The result is a cycle-organized chain, as shown in Figure~\ref{fig:overview}. Algorithm~\ref{alg:segraph} constructs $G$ from a parsed RTL design $D$ by mapping signals to $V_{\text{sig}}$, update logic to $V_{\text{stmt}}$ (marked combinational or clocked), and read/write dependencies to $E$.

\begin{algorithm}[t]
\caption{Construction of SeGraph.}
\label{alg:segraph}
\begin{algorithmic}[1]
\Require RTL design $D$
\Ensure $G = (V_{\text{sig}} \cup V_{\text{stmt}}, E)$
\State Parse continuous assignments, procedural assignments, and control structures in $D$
\State $V_{\text{sig}} \gets \{ x \mid x \text{ is a signal in } D \}$
\State $V_{\text{stmt}} \gets \emptyset$, $E \gets \emptyset$
\For{each combinational or clocked process $P$ in $D$}
    \For{each signal $y$ updated by $P$}
        \State Create a statement node $b$ from $y$'s update logic and enclosing \verb|if|/\verb|case| conditions
        \State Mark $b$ as combinational or clocked
        \State $V_{\text{stmt}} \gets V_{\text{stmt}} \cup \{b\}$
        \State $X \gets \{ x \in V_{\text{sig}} \mid x \text{ is read by } b \}$
        \State $E \gets E \cup \{ (x, b) \mid x \in X \} \cup \{ (b, y) \}$
    \EndFor
\EndFor
\State \Return $G$
\end{algorithmic}
\end{algorithm}

\subsection{Discussion}

The design of SeQuery and SeGraph reflects a tradeoff between tool completeness and LLM usability.
Formal completeness requires covering every temporal expression, but the resulting syntax and semantics are difficult for LLMs to use reliably.
Conversely, optimizing purely for ease of use risks leaving important verification scenarios uncovered.
SVA illustrates the former problem: although it can express arbitrary temporal properties, LLMs are prone to producing erroneous SVA assertions~\cite{assertion-generation,assertionbench}, which limits its practical value in agentic workflows.
Our design navigates this tension through the following choices:

\begin{itemize}
  \item SeQuery builds on SQL syntax and introduces only one additional construct beyond standard select-where: the $s@k$ relative-cycle offset. This keeps the learning cost near zero for LLMs while enabling cross-cycle queries. As Table~\ref{tab:sequery-examples} shows, this minimal design already covers a range of common sequential verification needs.

  \item SeGraph unifies signals, assignment statements, and conditional branches into a single dependency graph, capturing the dominant causal structures in RTL. An LLM agent specifies only the signals of interest and receives a subgraph organized by clock cycle, which makes cross-cycle value propagation visually interpretable without reasoning over raw source code.
\end{itemize}

This design does leave some scenarios uncovered, such as multi-clock-domain designs.
Overall, it represents a deliberate choice to favor LLM accessibility at the cost of some completeness, with the aim of enabling agents to obtain stable, actionable sequential feedback during iterative debugging.

\section{Evaluation}

\begin{table}[t]
    \centering
    \small
    \caption{Distribution of benchmark cases across categories.}
    \label{tab:benchmark-categories}
    \begin{tabular}{lr@{\hspace{1.5em}}lr}
        \toprule
        \textbf{Category} & \textbf{\#} & \textbf{Category} & \textbf{\#} \\
        \midrule
        Digital Signal Processing & 55 & Memory \& Bus & 30 \\
        Machine Learning & 57 & Control-Driven & 32 \\
        Accelerator & 58 & Heavy IP Integration & 24 \\
        \bottomrule
    \end{tabular}
\end{table}


\begin{figure*}[t]
    \centering
    \includegraphics[width=\textwidth]{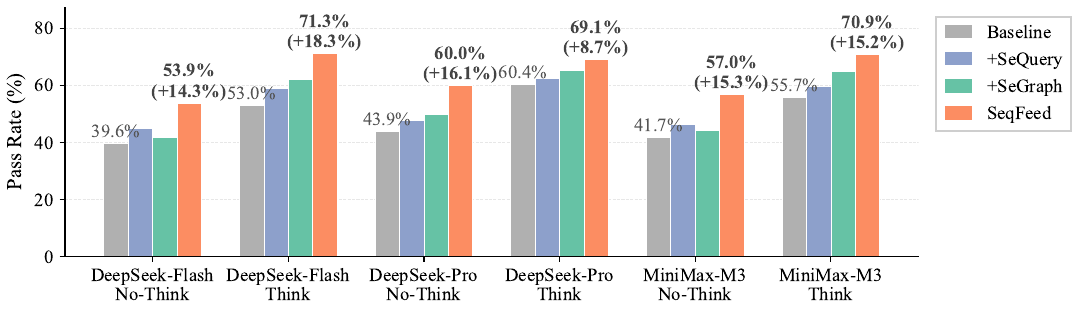}
    \caption{Pass rate across model configurations.}
    \label{fig:pass-rate}
\end{figure*}

\begin{figure}[t]
    \centering
    \includegraphics[width=\columnwidth]{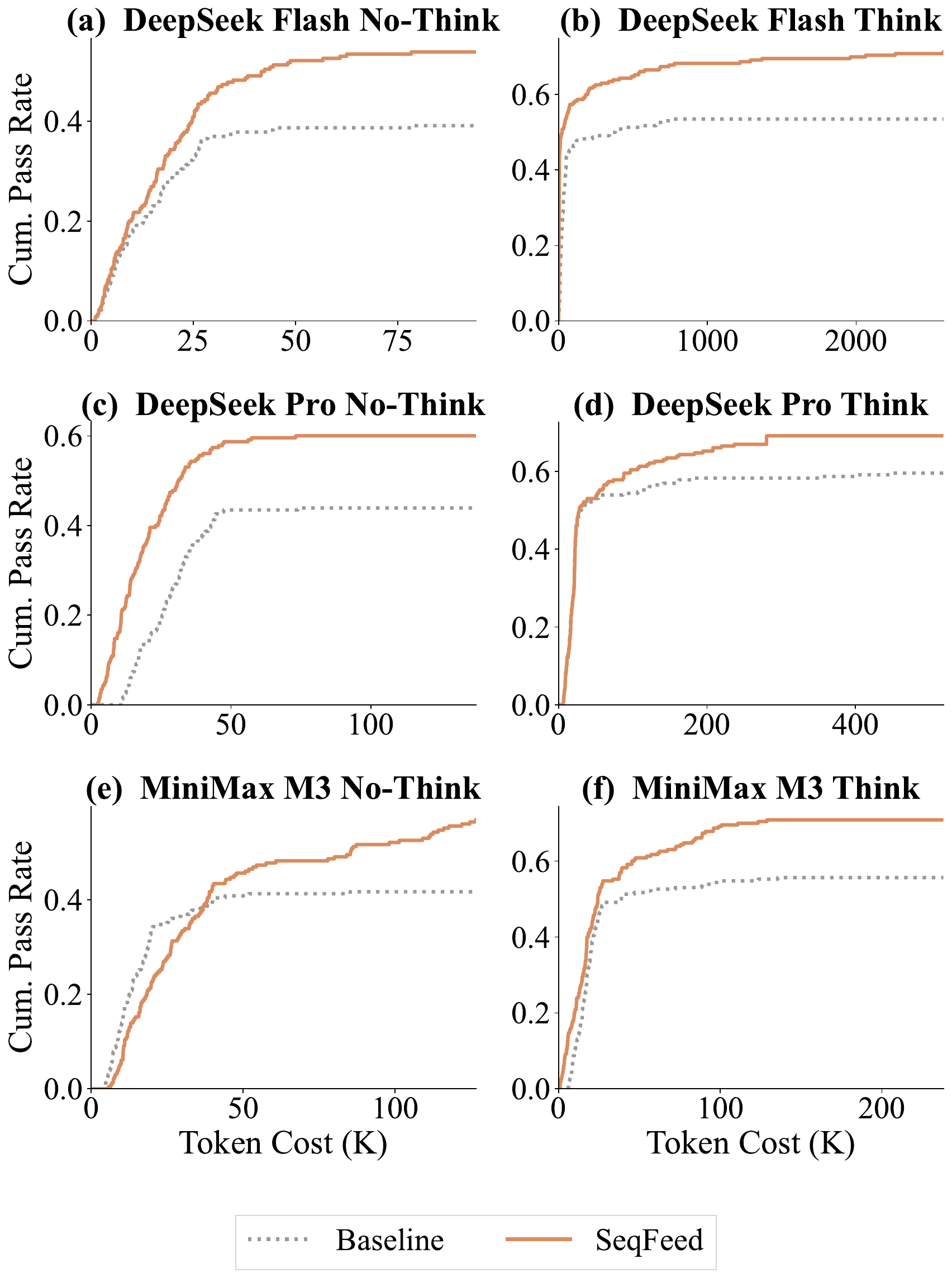}
    \caption{Cumulative pass rate against cumulative token cost.}
    \label{fig:efficiency}
\end{figure}

\begin{figure}[t]
    \centering
    \includegraphics[width=\columnwidth]{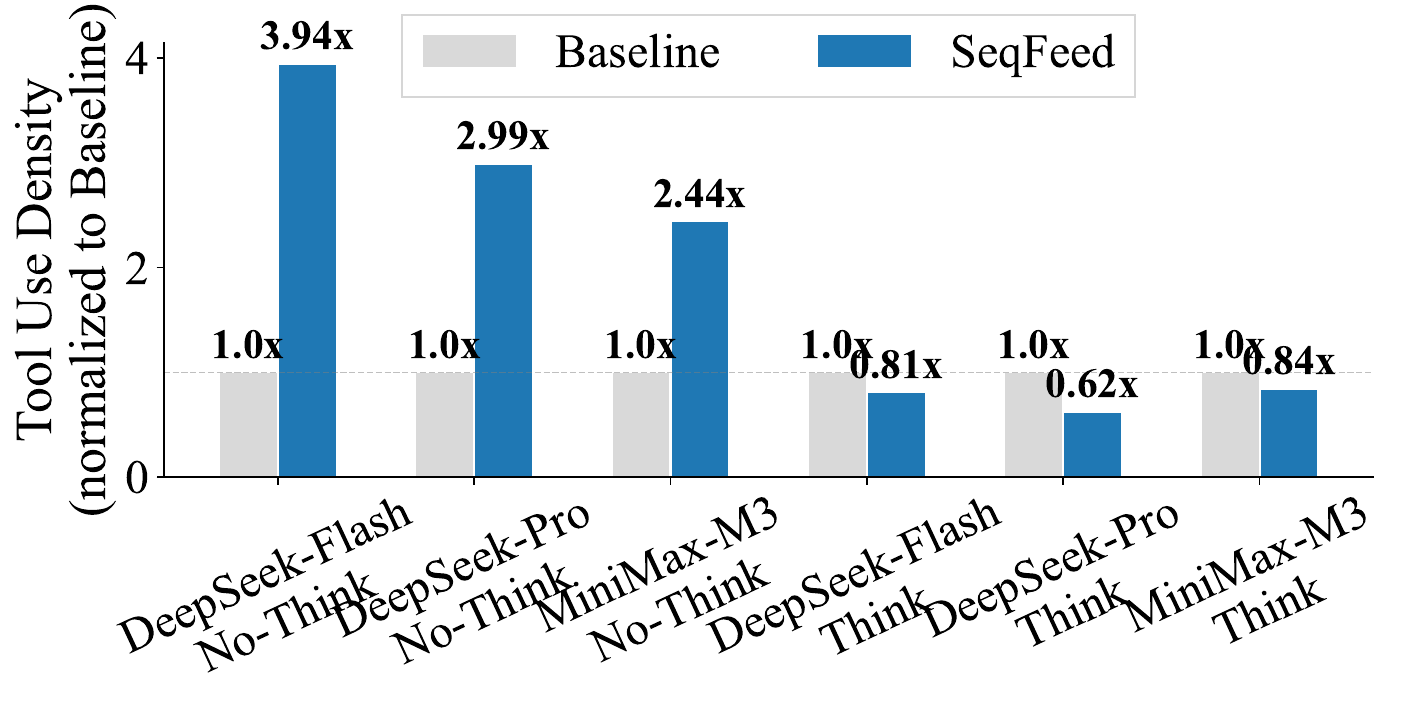}
    \caption{Tool use density across model configurations.}
    \label{fig:tool-use-density}
\end{figure}

\subsection{Experimental Setup}

Our benchmark combines cases from existing RTL generation benchmarks, including CVDP~\cite{pinckney2025cvdp} and RTLLM~\cite{lu2024rtllm}, with hardware design cases from recent open-source work~\cite{uvllm,forgebench}.
We exclude purely combinational circuits and toy cases that do not arise in real design practice, keeping the benchmark representative of practical hardware design.
Table~\ref{tab:benchmark-categories} shows the distribution of the 256 cases across six categories.

We evaluate on six model settings: DeepSeek Flash (no-think and think modes), DeepSeek Pro (no-think and think modes), and MiniMax M3 (no-think and think modes).
For each model and configuration, we run all 256 benchmark cases.
To isolate the contribution of each component, we compare four configurations: (1) baseline, where the agent receives only simulation error messages and neither SeQuery nor SeGraph are available; (2) SeQuery only; (3) SeGraph only; (4) SeqFeed, where both components are available. All other settings remain identical.
Agentic RTL generation is a long-horizon task spanning multiple rounds of iteration, each consuming tokens for context and generation.
We compare configurations by token cost: under a fixed token budget per case, we record whether the agent produces a passing design.

We use OpenCode~\cite{opencode} as the agent framework, with up to 65 steps per case.
Simulation uses Icarus Verilog, and test frameworks are built with cocotb~\cite{cocotb}.
Each agent run executes in an isolated Docker container with access only to the design specification and the test framework.
A case passes if the final RTL produced within the step budget passes all cocotb tests.

\subsection{Main Results}
\label{sec:main-results}

Figure~\ref{fig:pass-rate} reports the pass rate for each model setting under the four configurations.
SeqFeed improves pass rate over Baseline across all six model settings.
The gain ranges from 8.7 percentage points on DeepSeek Pro think (from 60.4\% to 69.1\%) to 18.3 percentage points on DeepSeek Flash think (from 53.0\% to 71.3\%).
In absolute terms, SeqFeed achieves the highest pass rate on DeepSeek Flash think (71.3\%), followed by MiniMax M3 think (70.9\%) and DeepSeek Pro think (69.1\%).
The improvement persists across both Flash and Pro tiers and across both think and no-think modes. Structured sequential feedback provides gains independent of model capacity and reasoning mode: even the strongest baseline, DeepSeek Pro think at 60.4\%, benefits from SeqFeed.
Ablation study shows that both SeQuery and SeGraph independently contribute to the pass rate gain.
With SeQuery alone, pass rate rises above Baseline by 2.2 to 6.1 percentage points across the six model settings.
With SeGraph alone, the gain ranges from 2.1 to 9.2 percentage points.
The full SeqFeed configuration achieves the highest pass rate on every model setting, reaching 71.3\% on DeepSeek Flash think.
On every model setting, the SeqFeed gain over Baseline (8.7--18.3 percentage points) exceeds the sum of the SeQuery-only and SeGraph-only gains, indicating that the two components reinforce each other.


\subsection{Token Utilization}
\label{sec:token-utilization}

Figure~\ref{fig:efficiency} plots cumulative pass rate against cumulative token cost. A flat segment indicates that further token spending produces no new passes.
Across all six settings, Baseline's gains are concentrated in the first 100K tokens and then nearly stop. DeepSeek Flash Think is representative: Baseline gains 46.5 percentage points in the first 100K, then only 6.9 points across the remaining 679K. SeqFeed gains 57.8 points in the first 100K and continues at 13.5 points across the subsequent range, roughly double Baseline's post-100K pace. The flat region of each Baseline curve measures debugging iterations that consume tokens without converging.

The MiniMax M3 No-Think curves make the trade-off explicit. Baseline leads early (13.5\% at 10K tokens against SeqFeed's 6.1\%) but stops gaining after 30K, finishing at 41.7\%. SeqFeed overtakes and reaches 57.0\%. This is the only setting where Baseline starts ahead; it is also where Baseline's productive range is narrowest. Structured feedback imposes a measurable upfront cost, but that cost buys access to cases that simple retry strategies cannot solve.


\subsection{Tool Use Density}
\label{sec:tool-use-density}

To characterize how the agent allocates its token budget between reasoning and action, we define \textit{tool use density} as the number of tool invocations per output token:
\[
\text{tool use density} = \frac{\#\text{tool invocations}}{\#\text{output tokens}} .
\]
We count only tool calls that advance the iterative RTL generation loop: editing the design, running simulation, and querying SeQuery or SeGraph. Calls unrelated to iteration (e.g., reading the specification or listing files) are excluded, as they do not reflect the debugging strategy.
Figure~\ref{fig:tool-use-density} reports tool use density across the six model settings, normalized so that each model's Baseline equals 1.0. A value above 1.0 means the configuration calls tools at a higher rate per token than Baseline; a value below 1.0 means it calls them at a lower rate.
The results reveal a sharp split by reasoning mode:

\begin{itemize}
    \item \textit{No-Think models.}
    On these three settings, SeqFeed raises tool use density substantially, to 2.4--3.9$\times$ the Baseline rate. Without a thinking mode, these models debug primarily through text: they generate hypotheses in natural language, edit the design, and simulate. Tool calls are sparse because the available tools do not provide waveform feedback; the agent fills the gap with internal reasoning. SeqFeed provides concrete entry points for action: SeQuery tests a hypothesis about a specific cycle, and SeGraph traces the dependency chain. The agent responds by invoking tools more frequently per token, replacing speculative reasoning with evidence-seeking queries.

    \item \textit{Think models.}
    On all three think settings, SeqFeed \emph{lowers} tool use density to 0.62--0.84$\times$ Baseline. With thinking mode, these models already reason about their debugging strategy in structured thought tokens. In Baseline, they make tool calls, but each call yields limited information: without waveform access, the agent probes repeatedly to narrow down a bug. SeqFeed makes each call more informative. A single SeQuery can confirm or refute a hypothesis that would otherwise require several speculative edit and simulate cycles. The per-token tool call rate drops because each call yields more information per query. DeepSeek Pro Think illustrates the shift: pass rate rises from 60.4\% to 69.1\% while tool calls per token fall to 0.62$\times$ Baseline.
\end{itemize}



\section{Related Work}

\paragraph{LLM-Aided Hardware Design}
Hardware design translates functional specifications into RTL implementations and is central to modern chip development. LLMs have been applied across the EDA pipeline, including RTL generation, verification, and optimization~\cite{zhong2024llm4eda,pan2025survey}. The agentic paradigm has been introduced to RTL tasks, where LLM agents autonomously invoke tools and iteratively refine their output~\cite{agentic_optimize,specloop}. For RTL generation, prior work uses iterative compile-and-simulate workflows~\cite{thakur2023autochip,ranga2024rtlagent,zhao2024mage}, fine-tunes LLMs on Verilog datasets~\cite{thakur2024verigen,chang2024data}, and introduces benchmarks for evaluating generated RTL~\cite{lu2024rtllm,pinckney2025cvdp}. Other efforts target Chisel generation~\cite{zhao2025codev,niu2025rechisel} or transpile C/Python into synthesizable Verilog~\cite{liao2024llms4hls,firouzi2024llmaid}. 

\paragraph{Sequential Feedback}
Effective feedback is essential for LLM agents to iteratively refine their outputs~\cite{gou2024critic,madaan2023selfrefine}. Software engineering agents commonly rely on compiler diagnostics and unit-test outcomes to identify program-level errors~\cite{chen2024selfdebugging,shinn2023reflexion}. RTL agents similarly use compiler errors, simulation results, EDA reports, and waveform traces~\cite{tsai2024rtlfixer,xu2024meic,ho2025verilogcoder,blocklove2025eda}. In particular, systems such as VerilogCoder~\cite{ho2025verilogcoder} and BlockLove~\cite{blocklove2025eda} show that waveform slices and timing diagrams can improve RTL correction by exposing cycle-level behavior. However, existing waveform feedback is typically preselected and static, preventing agents from querying transitions of interest, adjusting the temporal scope, or following signal dependencies across cycles. This limitation is compounded by evidence that LLMs continue to struggle with state transitions and cross-cycle dependencies~\cite{assertion-generation,assertionbench}. Providing sequential feedback that agents can query and explore according to their evolving hypotheses therefore remains an open challenge.

\section{Conclusion}

We identified the absence of effective sequential feedback as a central bottleneck in agentic RTL generation. By examining how human engineers interpret sequential behavior, we derived three requirements for such feedback: event-addressability, dependency-traceability, and incremental expandability. To meet these requirements, we built SeqFeed, which combines SeQuery, a waveform query language, with SeGraph, a cycle-organized dependency graph. Experiments across six LLM configurations show that SeqFeed improves pass rate and token efficiency, with the two components independently effective and complementary. Analysis of agent reasoning further indicates a shift from speculation-driven conjecture to evidence-driven diagnosis.

\bibliography{ref}

@article{wang2024survey,
  title     = {A Survey on Large Language Model Based Autonomous Agents},
  author    = {Lei Wang and Chen Ma and Xueyang Feng and Zeyu Zhang and Hao Yang and Jingsen Zhang and Zhiyuan Chen and Jiakai Tang and Xu Chen and Yankai Lin and Wayne Xin Zhao and Zhewei Wei and Ji-Rong Wen},
  journal   = {Frontiers of Computer Science},
  volume    = {18},
  number    = {6},
  year      = {2024},
  doi       = {10.1007/s11704-024-40231-1},
}

@misc{yao2025arspautomatedrepairverilog,
      title={ARSP: Automated Repair of Verilog Designs via Semantic Partitioning}, 
      author={Bingkun Yao and Ning Wang and Xiangfeng Liu and Yuxin Du and Yuchen Hu and Hong Gao and Zhe Jiang and Nan Guan},
      year={2025},
      eprint={2508.16517},
      archivePrefix={arXiv},
      primaryClass={cs.SE},
      url={https://arxiv.org/abs/2508.16517}, 
}

@inproceedings{
yao2023react,
title={ReAct: Synergizing Reasoning and Acting in Language Models},
author={Shunyu Yao and Jeffrey Zhao and Dian Yu and Nan Du and Izhak Shafran and Karthik R Narasimhan and Yuan Cao},
booktitle={The Eleventh International Conference on Learning Representations },
year={2023},
url={https://openreview.net/forum?id=WE_vluYUL-X}
}

@inproceedings{madaan2023selfrefine,
author = {Madaan, Aman and Tandon, Niket and Gupta, Prakhar and Hallinan, Skyler and Gao, Luyu and Wiegreffe, Sarah and Alon, Uri and Dziri, Nouha and Prabhumoye, Shrimai and Yang, Yiming and Gupta, Shashank and Majumder, Bodhisattwa Prasad and Hermann, Katherine and Welleck, Sean and Yazdanbakhsh, Amir and Clark, Peter},
title = {SELF-REFINE: iterative refinement with self-feedback},
year = {2023},
publisher = {Curran Associates Inc.},
address = {Red Hook, NY, USA},
booktitle = {Proceedings of the 37th International Conference on Neural Information Processing Systems},
articleno = {2019},
numpages = {61},
location = {New Orleans, LA, USA},
series = {NIPS '23}
}

@inproceedings{
gou2024critic,
title={{CRITIC}: Large Language Models Can Self-Correct with Tool-Interactive Critiquing},
author={Zhibin Gou and Zhihong Shao and Yeyun Gong and yelong shen and Yujiu Yang and Nan Duan and Weizhu Chen},
booktitle={The Twelfth International Conference on Learning Representations},
year={2024},
url={https://openreview.net/forum?id=Sx038qxjek}
}

@INPROCEEDINGS{liu2023verilogeval,
  author={Liu, Mingjie and Pinckney, Nathaniel and Khailany, Brucek and Ren, Haoxing},
  booktitle={2023 IEEE/ACM International Conference on Computer Aided Design (ICCAD)}, 
  title={Invited Paper: VerilogEval: Evaluating Large Language Models for Verilog Code Generation}, 
  year={2023},
  volume={},
  number={},
  pages={1-8},
  doi={10.1109/ICCAD57390.2023.10323812}}

@inproceedings{tsai2024rtlfixer,
author = {Tsai, Yunda and Liu, Mingjie and Ren, Haoxing},
title = {RTLFixer: Automatically Fixing RTL Syntax Errors with Large Language Model},
year = {2024},
isbn = {9798400706011},
publisher = {Association for Computing Machinery},
address = {New York, NY, USA},
url = {https://doi.org/10.1145/3649329.3657353},
doi = {10.1145/3649329.3657353},
booktitle = {Proceedings of the 61st ACM/IEEE Design Automation Conference},
articleno = {53},
numpages = {6},
location = {San Francisco, CA, USA},
series = {DAC '24}
}

@inproceedings{chang2024data,
author = {Chang, Kaiyan and Wang, Kun and Yang, Nan and Wang, Ying and Jin, Dantong and Zhu, Wenlong and Chen, Zhirong and Li, Cangyuan and Yan, Hao and Zhou, Yunhao and Zhao, Zhuoliang and Cheng, Yuan and Pan, Yudong and Liu, Yiqi and Wang, Mengdi and Liang, Shengwen and Han, Yinhe and Li, Huawei and Li, Xiaowei},
title = {Data is all you need:  Finetuning LLMs for Chip Design via an Automated design-data augmentation framework},
year = {2024},
isbn = {9798400706011},
publisher = {Association for Computing Machinery},
address = {New York, NY, USA},
url = {https://doi.org/10.1145/3649329.3657356},
doi = {10.1145/3649329.3657356},
booktitle = {Proceedings of the 61st ACM/IEEE Design Automation Conference},
articleno = {60},
numpages = {6},
location = {San Francisco, CA, USA},
series = {DAC '24}
}

@Article{abdollahi2024hardware,
AUTHOR = {Abdollahi, Meisam and Yeganli, Seyedeh Faegheh and Baharloo, Mohammad (Amir) and Baniasadi, Amirali},
TITLE = {Hardware Design and Verification with Large Language Models: A Scoping Review, Challenges, and Open Issues},
JOURNAL = {Electronics},
VOLUME = {14},
YEAR = {2025},
NUMBER = {1},
ARTICLE-NUMBER = {120},
URL = {https://www.mdpi.com/2079-9292/14/1/120},
ISSN = {2079-9292},
DOI = {10.3390/electronics14010120}
}

@inproceedings{xu2024meic,
author = {Xu, Ke and Sun, Jialin and Hu, Yuchen and Fang, Xinwei and Shan, Weiwei and Wang, Xi and Jiang, Zhe},
title = {MEIC: Re-thinking RTL Debug Automation using LLMs},
year = {2025},
isbn = {9798400710773},
publisher = {Association for Computing Machinery},
address = {New York, NY, USA},
url = {https://doi.org/10.1145/3676536.3676801},
doi = {10.1145/3676536.3676801},
booktitle = {Proceedings of the 43rd IEEE/ACM International Conference on Computer-Aided Design},
articleno = {100},
numpages = {9},
location = {Newark Liberty International Airport Marriott, New York, NY, USA},
series = {ICCAD '24}
}

@misc{zhong2024llm4eda,
      title={LLM4EDA: Emerging Progress in Large Language Models for Electronic Design Automation}, 
      author={Ruizhe Zhong and Xingbo Du and Shixiong Kai and Zhentao Tang and Siyuan Xu and Hui-Ling Zhen and Jianye Hao and Qiang Xu and Mingxuan Yuan and Junchi Yan},
      year={2023},
      eprint={2401.12224},
      archivePrefix={arXiv},
      primaryClass={cs.AR},
      url={https://arxiv.org/abs/2401.12224}, 
}

@article{pan2025survey,
author = {Pan, Jingyu and Zhou, Guanglei and Chang, Chen-Chia and Jacobson, Isaac and Hu, Jiang and Chen, Yiran},
title = {A Survey of Research in Large Language Models for Electronic Design Automation},
year = {2025},
issue_date = {May 2025},
publisher = {Association for Computing Machinery},
address = {New York, NY, USA},
volume = {30},
number = {3},
issn = {1084-4309},
url = {https://doi.org/10.1145/3715324},
doi = {10.1145/3715324},
journal = {ACM Trans. Des. Autom. Electron. Syst.},
month = feb,
articleno = {34},
numpages = {21}
}

@misc{thakur2023autochip,
      title={AutoChip: Automating HDL Generation Using LLM Feedback}, 
      author={Shailja Thakur and Jason Blocklove and Hammond Pearce and Benjamin Tan and Siddharth Garg and Ramesh Karri},
      year={2024},
      eprint={2311.04887},
      archivePrefix={arXiv},
      primaryClass={cs.PL},
      url={https://arxiv.org/abs/2311.04887}, 
}

@INPROCEEDINGS{ranga2024rtlagent,
  author={Ranga, Sriram and Mao, Rui and Bhattacharjee, Debjyoti and Cambria, Erik and Chattopadhyay, Anupam},
  booktitle={2024 IEEE 33rd Asian Test Symposium (ATS)}, 
  title={RTL Agent: An Agent-Based Approach for Functionally Correct HDL Generation via LLMs}, 
  year={2024},
  volume={},
  number={},
  pages={1-6},
  doi={10.1109/ATS64447.2024.10915277}}

@article{thakur2024verigen,
author = {Thakur, Shailja and Ahmad, Baleegh and Pearce, Hammond and Tan, Benjamin and Dolan-Gavitt, Brendan and Karri, Ramesh and Garg, Siddharth},
title = {VeriGen: A Large Language Model for Verilog Code Generation},
year = {2024},
issue_date = {May 2024},
publisher = {Association for Computing Machinery},
address = {New York, NY, USA},
volume = {29},
number = {3},
issn = {1084-4309},
url = {https://doi.org/10.1145/3643681},
doi = {10.1145/3643681},
journal = {ACM Trans. Des. Autom. Electron. Syst.},
month = apr,
articleno = {46},
numpages = {31}
}

@INPROCEEDINGS{lu2024rtllm,
  author={Lu, Yao and Liu, Shang and Zhang, Qijun and Xie, Zhiyao},
  booktitle={2024 29th Asia and South Pacific Design Automation Conference (ASP-DAC)}, 
  title={RTLLM: An Open-Source Benchmark for Design RTL Generation with Large Language Model}, 
  year={2024},
  volume={},
  number={},
  pages={722-727},
  doi={10.1109/ASP-DAC58780.2024.10473904}}

@misc{pinckney2025cvdp,
      title={Comprehensive Verilog Design Problems: A Next-Generation Benchmark Dataset for Evaluating Large Language Models and Agents on RTL Design and Verification}, 
      author={Nathaniel Pinckney and Chenhui Deng and Chia-Tung Ho and Yun-Da Tsai and Mingjie Liu and Wenfei Zhou and Brucek Khailany and Haoxing Ren},
      year={2025},
      eprint={2506.14074},
      archivePrefix={arXiv},
      primaryClass={cs.LG},
      url={https://arxiv.org/abs/2506.14074}, 
}

@ARTICLE{zhao2025codev,
  author={Zhao, Yang and Huang, Di and Li, Chongxiao and Jin, Pengwei and Song, Muxin and Xu, Yinan and Nan, Ziyuan and Gao, Mingju and Ma, Tianyun and Qi, Lei and Pan, Yansong and Zhang, Zhenxing and Zhang, Rui and Zhang, Xishan and Du, Zidong and Guo, Qi and Hu, Xing},
  journal={IEEE Transactions on Computer-Aided Design of Integrated Circuits and Systems}, 
  title={CodeV: Empowering LLMs With HDL Generation Through Multilevel Summarization}, 
  year={2026},
  volume={45},
  number={4},
  pages={1893-1906},
  doi={10.1109/TCAD.2025.3604320}}

@INPROCEEDINGS{niu2025rechisel,
  author={Niu, Juxin and Liu, Xiangfeng and Niu, Dan and Wang, Xi and Jiang, Zhe and Guan, Nan},
  booktitle={2025 62nd ACM/IEEE Design Automation Conference (DAC)}, 
  title={ReChisel: Effective Automatic Chisel Code Generation by LLM with Reflection}, 
  year={2025},
  volume={},
  number={},
  pages={1-7},
  doi={10.1109/DAC63849.2025.11132940}}

@inproceedings{liao2024llms4hls,
author = {Liao, Yuchao and Adegbija, Tosiron and Lysecky, Roman},
title = {Are LLMs Any Good for High-Level Synthesis?},
year = {2025},
isbn = {9798400710773},
publisher = {Association for Computing Machinery},
address = {New York, NY, USA},
url = {https://doi.org/10.1145/3676536.3699507},
doi = {10.1145/3676536.3699507},
booktitle = {Proceedings of the 43rd IEEE/ACM International Conference on Computer-Aided Design},
articleno = {29},
numpages = {8},
location = {Newark Liberty International Airport Marriott, New York, NY, USA},
series = {ICCAD '24}
}

@INPROCEEDINGS{zhao2024mage,
  author={Zhao, Yujie and Zhang, Hejia and Huang, Hanxian and Yu, Zhongming and Zhao, Jishen},
  booktitle={2025 62nd ACM/IEEE Design Automation Conference (DAC)}, 
  title={MAGE: A Multi-Agent Engine for Automated RTL Code Generation}, 
  year={2025},
  volume={},
  number={},
  pages={1-7},
  doi={10.1109/DAC63849.2025.11133191}}

@inproceedings{firouzi2024llmaid,
author = {Firouzi, Farshad and Nakkilla, Sri Sai Rakesh and Fu, Chenghao and Banerjee, Sanmitra and Talukdar, Jonti and Chakrabarty, Krishnendu},
title = {LLM-AID: Leveraging Large Language Models for Rapid Domain-Specific Accelerator Development},
year = {2025},
isbn = {9798400710773},
publisher = {Association for Computing Machinery},
address = {New York, NY, USA},
url = {https://doi.org/10.1145/3676536.3697135},
doi = {10.1145/3676536.3697135},
booktitle = {Proceedings of the 43rd IEEE/ACM International Conference on Computer-Aided Design},
articleno = {25},
numpages = {9},
location = {Newark Liberty International Airport Marriott, New York, NY, USA},
series = {ICCAD '24}
}

@inproceedings{
chen2024selfdebugging,
title={Teaching Large Language Models to Self-Debug},
author={Xinyun Chen and Maxwell Lin and Nathanael Sch{\"a}rli and Denny Zhou},
booktitle={The Twelfth International Conference on Learning Representations},
year={2024},
url={https://openreview.net/forum?id=KuPixIqPiq}
}

@inproceedings{shinn2023reflexion,
author = {Shinn, Noah and Cassano, Federico and Gopinath, Ashwin and Narasimhan, Karthik and Yao, Shunyu},
title = {Reflexion: language agents with verbal reinforcement learning},
year = {2023},
publisher = {Curran Associates Inc.},
address = {Red Hook, NY, USA},
booktitle = {Proceedings of the 37th International Conference on Neural Information Processing Systems},
articleno = {377},
numpages = {19},
location = {New Orleans, LA, USA},
series = {NIPS '23}
}

@inproceedings{ho2025verilogcoder,
author = {Ho, Chia-Tung and Ren, Haoxing and Khailany, Brucek},
title = {VerilogCoder: autonomous verilog coding agents with graph-based planning and abstract syntax tree (AST)-based waveform tracing tool},
year = {2025},
isbn = {978-1-57735-897-8},
publisher = {AAAI Press},
url = {https://doi.org/10.1609/aaai.v39i1.32007},
doi = {10.1609/aaai.v39i1.32007},
articleno = {34},
numpages = {8},
booktitle = {Proceedings of the Thirty-Ninth AAAI Conference on Artificial Intelligence and Thirty-Seventh Conference on Innovative Applications of Artificial Intelligence and Fifteenth Symposium on Educational Advances in Artificial Intelligence},
series = {AAAI'25/IAAI'25/EAAI'25}
}

@article{blocklove2025eda,
author = {Blocklove, Jason and Thakur, Shailja and Tan, Benjamin and Pearce, Hammond and Garg, Siddharth and Karri, Ramesh},
title = {Automatically Improving LLM-based Verilog Generation using EDA Tool Feedback},
year = {2025},
issue_date = {November 2025},
publisher = {Association for Computing Machinery},
address = {New York, NY, USA},
volume = {30},
number = {6},
issn = {1084-4309},
url = {https://doi.org/10.1145/3723876},
doi = {10.1145/3723876},
journal = {ACM Trans. Des. Autom. Electron. Syst.},
month = oct,
articleno = {100},
numpages = {26}
}

@inproceedings{assertionbench,
    title     = {{A}ssertion{B}ench: A Benchmark to Evaluate Large-Language Models for Assertion Generation},
    author = {Pulavarthi, Vaishnavi  and  Nandal, Deeksha  and  Dan, Soham  andPal, Debjit},
    booktitle = "Findings of the Association for Computational Linguistics: NAACL 2025",
    month = {apr},
    year = {2025},
    address = {Albuquerque, New Mexico},
    publisher = {Association for Computational Linguistics},
    url = {https://aclanthology.org/2025.findings-naacl.449/},
    doi = {10.18653/v1/2025.findings-naacl.449},
    pages = {8073--8080},
    ISBN = {979-8-89176-195-7}
}

@INPROCEEDINGS{assertion-generation,
  author={Pulavarthi, Vaishnavi and Nandal, Deeksha and Dan, Soham and Pal, Debjit},
  booktitle={2025 Design, Automation \& Test in Europe Conference (DATE)}, 
  title={Are LLMs Ready for Practical Adoption for Assertion Generation?}, 
  year={2025},
  volume={},
  number={},
  pages={1-7},
  doi={10.23919/DATE64628.2025.10992817}}

@misc{opencode,
  title        = {OpenCode: The Open Source AI Coding Agent},
  author       = {{Anomaly}},
  year         = {2025},
  howpublished = {\url{https://github.com/anomalyco/opencode}},
  note         = {Accessed: 2026-07-27},
}

@misc{cocotb,
  title        = {cocotb: Python-based Chip RTL Verification},
  author       = {{cocotb contributors}},
  year         = {2025},
  howpublished = {\url{https://github.com/cocotb/cocotb}},
  note         = {Accessed: 2026-07-27},
}

@INPROCEEDINGS{uvllm,
  author={Hu, Yuchen and Ye, Junhao and Xu, Ke and Sun, Jialin and Zhang, Shiyue and Jiao, Xinyao and Pan, Dingrong and Zhou, Jie and Wang, Ning and Shan, Weiwei and Fang, Xinwei and Wang, Xi and Guan, Nan and Jiang, Zhe},
  booktitle={2025 62nd ACM/IEEE Design Automation Conference (DAC)}, 
  title={UVLLM: An Automated Universal RTL Verification Framework using LLMs}, 
  year={2025},
  volume={},
  number={},
  pages={1-7},
  doi={10.1109/DAC63849.2025.11435108}}

@misc{agentic_optimize,
      title={Dr. RTL: Autonomous Agentic RTL Optimization through Tool-Grounded Self-Improvement}, 
      author={Wenji Fang and Yao Lu and Shang Liu and Jing Wang and Ziyan Guo and Junxian He and Fengbin Tu and Zhiyao Xie},
      year={2026},
      eprint={2604.14989},
      archivePrefix={arXiv},
      primaryClass={cs.AI},
      url={https://arxiv.org/abs/2604.14989}, 
}

@misc{specloop,
      title={SpecLoop: An Agentic RTL-to-Specification Framework with Formal Verification Feedback Loop}, 
      author={Fu-Chieh Chang and Yu-Hsin Yang and Hung-Ming Huang and Yun-Chia Hsu and Yin-Yu Lin and Ming-Fang Tsai and Chun-Chih Yang and Pei-Yuan Wu},
      year={2026},
      eprint={2603.02895},
      archivePrefix={arXiv},
      primaryClass={cs.AR},
      url={https://arxiv.org/abs/2603.02895}, 
}

@misc{forgebench,
      title={ForgeBench: A Machine Learning Benchmark Suite and Auto-Generation Framework for Next-Generation HLS Tools}, 
      author={Andy Wanna and Hanqiu Chen and Cong Hao},
      year={2025},
      eprint={2504.15185},
      archivePrefix={arXiv},
      primaryClass={cs.AR},
      url={https://arxiv.org/abs/2504.15185}, 
}
\end{document}